 \documentclass[12pt,preprint]{aastex}

\slugcomment{The Astrophysical Journal, in press}

\shorttitle{The Neupert Effect in an RS CVn Binary}
\shortauthors{G\"udel et al.}

\begin{document}

\def\parn{\par\noindent}

\title{Detection of the Neupert Effect in the Corona of an RS CVn Binary System
       by {\it XMM-Newton} and the VLA}  

\author{Manuel G\"udel, Marc Audard, Kester W. Smith\altaffilmark{1}}
\affil{Paul Scherrer Institut, W\"urenlingen und Villigen, 
                 CH-5232 Villigen PSI, Switzerland}
\altaffiltext{1}{Also Institut f\"ur Astronomie, ETH Zentrum, CH-8092 Z\"urich, Switzerland}
\email{guedel@astro.phys.ethz.ch, audard@astro.phys.ethz.ch, kester@astro.phys.ethz.ch}

\author{Ehud Behar}
\affil{Columbia Astrophysics Laboratory, Columbia University, 550 West 120th Street, New York,
       NY 10027, USA}
\email{behar@astro.columbia.edu}

\author{Anthony J. Beasley}
\affil{Owens Valley Radio Observatory, California Institute of Technology, 
         Big Pine, CA 93513, USA}
\email{tbeasley@ovro.caltech.edu}

\and

\author{Rolf Mewe}
\affil{SRON National Institute for Space Research, Sorbonnelaan 2, 
                3584 CA Utrecht, The Netherlands}
\email{r.mewe@sron.nl\\ {\ } \\ {\ } \\ {\ } \\ {\ } \\ {\ } \\ {\ } \\}

\begin{abstract}
The RS CVn-type binary $\sigma$ Geminorum was observed during a large,
long-duration flare simultaneously with {\it XMM-Newton} and the VLA. The light curves show a 
characteristic time dependence that is compatible with the Neupert effect observed in solar
flares: The time derivative of the X-ray light curve resembles the radio light curve.
This observation can be interpreted in terms of a standard flare scenario in which 
accelerated coronal electrons reach the chromosphere where they heat the cool plasma 
and induce chromospheric evaporation. Such a scenario can only hold if the 
amount of energy in the fast electrons is sufficient to explain the X-ray
radiative losses. We present a plausibility analysis 
that supports the  chromospheric evaporation model.
\end{abstract}


\section{Introduction}

There is compelling evidence that high-energy processes and high-energy particles
play a pivotal role in the energy release, energy transport, and plasma heating 
during solar flares (see review by \citealt{hudson95}). A standard scenario proposes 
that electrons (perhaps also ions) are accelerated in the corona in the course of magnetic 
reconnection. As the electrons travel along closed magnetic fields, those  with large 
pitch angles and sufficient energy  (typically several 100~keV) lose a small part of their 
energy as  gyrosynchrotron radio emission. The bulk kinetic energy of the accelerated 
electrons, however, is carried to the chromosphere where it is deposited by electron-ion
collisions. The collision of the beam with the dense plasma reveals itself by non-thermal 
hard X-ray radiation (HXR, typically between 10$-$100~keV) that, however, constitutes
only a small fraction ($\sim 10^{-5}$)  of the total energy loss. The bulk energy is
transformed into heat, producing an overpressure in the chromosphere as the gas cannot
radiate away the energy influx sufficiently rapidly. As a consequence, the gas 
evaporates explosively into the corona as a $\sim 10^7$~K plasma visible in X-rays during the
gradual phase of a solar flare \citep{dennis88}. The observed gyrosynchrotron radio
luminosity $L_{\mathrm{R}}$ and the hard X-ray luminosity $L_{\mathrm{HXR}}$ are, 
to first order, proportional to the instantaneous number of fast electrons 
and therefore to the power $\dot{E}$ injected into the system, while the slowly variable 
soft X-ray luminosity $L_{\mathrm{X}}$ is roughly proportional  to the accumulated 
total energy $E$ in the hot coronal flare plasma. One therefore expects that 
\begin{equation}\label{neupert}
{d\over dt} L_{\mathrm{X}}(t) \propto L_{\mathrm{HXR}}(t) \propto L_{\mathrm{R}}(t) ,
\end{equation}
a relation that is commonly known as the `Neupert Effect' \citep{neupert68,dennis93}. 
Although significant deviations from this scenario have been observed in solar flares 
(e.g., heating starting before any hard X-rays can be detected, or absence of one of the 
emission types discussed above), there is strong support for several features of this
model in the majority of solar flares \citep{dennis93}. For example, the coincidence to
within a fraction of a second of HXR brightenings at a pair of magnetic loop footpoints
that are separated by $\sim 10^9$~cm  requires non-thermal particle velocities \citep{sakao94}. 

It is thought that the same mechanisms should operate in stellar flares,
although flares on some classes of stars deviate considerably from the proposed
solar analogy. In particular, giant flares on RS CVn-type binaries may require 
mechanisms unknown on the Sun. RS CVn binaries, commonly consisting of a giant or
subgiant primary with a main-sequence or subgiant companion in a close orbit, are sources 
of luminous radio and X-ray emission \citep*{drake89}. Very long flare time scales
and radio source sizes of order of the intrabinary distance have been modeled
in terms of giant dipole-like magnetospheric structures \citep*{morris90, jones94} into 
which high-energy particles are injected from a flare site \citep{mutel85}, and where 
they lose most of their energy by radiation. Coordinated observations in X-rays and radio
are required to study the importance of high-energy electrons in the heating mechanism.
\citet{hawley95}
reported a Neupert effect-like behavior during a large flare on the dMe star AD Leo,
where optical (U band) and EUV emissions were used as proxies for the radiation from
high-energy electrons
(radio, HXR) and from the thermal plasma,  respectively. \citet{guedel96} discussed the
first stellar Neupert effect seen in the  radio and X-ray bands on the dM5.5e
binary UV Cet, finding similar timing and similar energy budgets as in solar
gradual (``type C'') events.


\section{Observations and Data Analysis}

$\sigma$ Gem is a close binary of the RS CVn type, consisting of a K1~III giant and a fainter
companion whose spectral classification is unclear \citep{strassmeier93}.
Its distance is $37.5\pm 1$~pc \citep{esa97}.
It is a luminous radio (log$L_{\mathrm{R}} \approx 15.40$, in erg~s$^{-1}$~Hz$^{-1}$;
\citealt{drake89}) and X-ray source (log$L_{\mathrm{X}} \approx 31.0\pm 0.2$, in 
erg~s$^{-1}$, \citealt{yi97} and references therein, 
all luminosities corrected to the Hipparcos distance), and
was monitored previously during flares or during quiescence at radio, EUV, and 
X-ray wavelengths \citep{engvold88, pallavicini85, singh87, drake89, schrijver95, yi97, osten99}. 
It is generally assumed that the coronal emission is largely related to the 
K1~III star.

We observed $\sigma$ Gem with {\it XMM-Newton} \citep{jansen01} between
2001 April 6, 16:46~UT $-$ April 7, 07:53~UT, using
all five X-ray detectors on board. Since here we are predominantly 
interested in the time behavior, we will present exclusively the
sensitive European Photon Imaging Camera (EPIC) PN light curve for the energy 
range [0.15, 10]~keV \citep{strueder01}.
The PN camera was operated in the small window mode given the expected brightness 
of the target (both at optical and at X-ray wavelengths). Background subtraction
is irrelevant for the very high source count
rate. The features in the light curve were
confirmed from the other detectors. The data were processed using the
standard SAS software (version of 2001 May). Dead-time corrections were applied  
(29\% of the counts are lost in the utilized mode). For most of the observing time
(April 6, 20:12~UT $-$ April 7, 07:58~UT), 
the Very Large Array (VLA) monitored the 6~cm flux (frequency $\nu = 5\times 10^9$~Hz) of $\sigma$ Gem in 
both left- and  right-hand polarizations over a bandwidth of 100~MHz. The calibration of 
the radio data followed standard procedures within the AIPS software. Individual 
maps were made for each observing scan ($\approx$12.5$^{\prime}$); the fluxes were read from
these maps using the JMFIT task.

\section{Results}

\begin{figure}
\epsscale{0.6}
\plotone{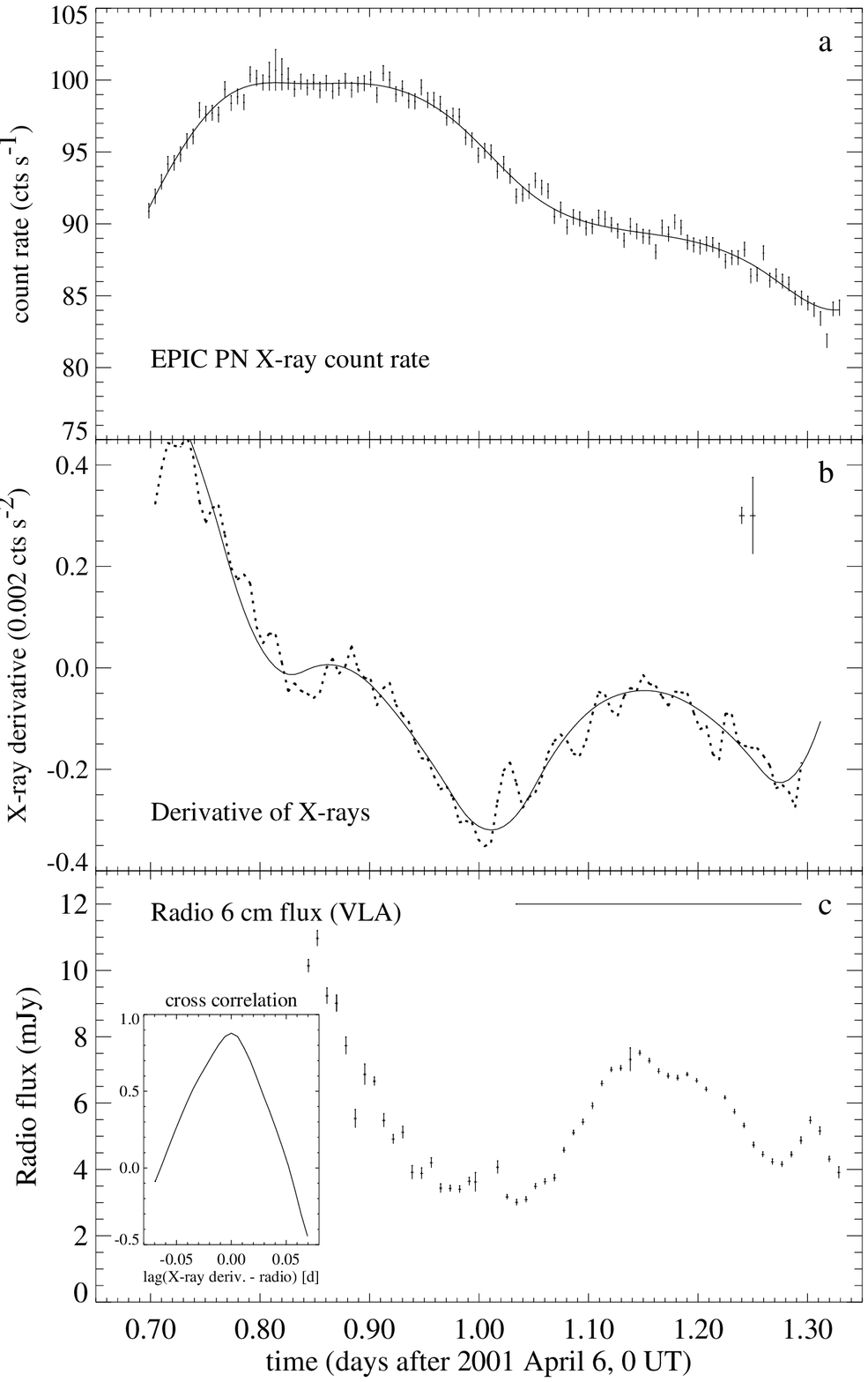}
\caption{Light curves of $\sigma$ Gem. {\bf a)} (top): {\it XMM-Newton}
      EPIC PN light curve, binned to 500~s and corrected for dead time. The smooth solid curve   
      has been obtained from  a Chebychev polynomial fit of order 9. 
      {\bf b)} (middle): The time derivative of the 
      smoothed X-ray light curve. The solid curve shows the time derivative of the Chebychev polynomial fit
      above; the dotted curve was derived from a boxcar-smoothed light curve (boxcar length = 11; 
      not shown).
      For the latter derivative, the larger error bar in the upper right corner illustrates the absolute 
      uncertainty for  any single derivative, whereas the smaller error bar indicates the relative 
      scatter between nearest neighbors and is smaller due to correlations introduced by the smoothing.
      {\bf c)} (bottom): The VLA 6~cm light curve, binned to $\sim$750~s (observing scan length). 
      The inset shows the cross-correlation function of the radio light curve and the X-ray time derivative,
      computed for the time interval marked with a horizontal bar above the second radio flare.
      \label{lightcurves}}
\end{figure}

{\it XMM-Newton} observed $\sigma$ Gem in the course of a large X-ray flare,
with a PN peak count rate of $\sim$100~cts~s$^{-1}$, corresponding to
a luminosity of $\sim 5\times 10^{31}$~erg~s$^{-1}$ or $\approx$5 times the usual 
quiescent X-ray emission level (as cited above). Figure~\ref{lightcurves}a (top panel) 
presents the PN light curve, binned to 500~s. Only the peak of this flare was observed. 
From the PN spectrum, the X-ray emission can be interpreted as being due to hot (several
keV), thermal plasma. The simultaneous radio
emission (bottom panel in Figure~\ref{lightcurves}) is strongly varying around
a few mJy ($=10^{-26}$~erg~cm$^{-2}~$s$^{-1}$~Hz$^{-1}$), typically 
on time scales of less than 5000~s. We will concentrate our discussion on
the second flare that is fully visible at radio wavelengths, i.e., the episode
between 1.03$-$1.32~d.

Motivated by equation~\ref{neupert}, we  smoothed the X-ray light curve, either
using a  boxcar of width 11 bins (equivalent to 5500~s, not shown),
or a Chebychev-polynomial fit of order 9 (Figure~\ref{lightcurves}a).  
We then  calculated the time derivative of each smoothed light curve. The results 
are illustrated in Figure~\ref{lightcurves}b (solid: for Chebychev fit; dotted: for 
boxcar-smoothed fit), and
are to be compared with the radio light curve in panel c. The  curves are
very similar, signifying a relation close to that described by equation~\ref{neupert}, i.e.,
a Neupert effect. To quantify the relative timing of the two curves, we 
computed the cross correlation function as follows:
\begin{equation}\label{cross}
P(\ell) = {m\over m-|\ell|}{\displaystyle{\sum_{k=1}^{m-\ell}}(R_k-\bar{R})(D_{k+\ell}-\bar{D}) \over
           \left(\left[\displaystyle{\sum_{k=1}^m}(R_k-\bar{R})^2\right]
	   \left[\displaystyle{\sum_{k=1}^m}(D_k-\bar{D})^2\right]\right)^{1/2}}
\end{equation}
for a time lag $\ell \ge 0$ in units of bins, where we used $m=46$ time bins of 500~s each
as shown  in Figure~\ref{lightcurves}b, for the time interval marked by a horizontal 
bar in  Figure~\ref{lightcurves}c. For $\ell < 0$, $D$ and $R$ are interchanged.
Here, $R_k$ is the radio flux and $D_k$ is the X-ray time derivative at the grid 
point $k$ (the boxcar-smoothed version was used; the result  using the Chebychev-smoothed 
version is in full agreement). The quantities $\bar{R}$ and $\bar{D}$ are the means
of the respective variables. The first factor on the right-hand side 
of equation~\ref{cross}  corrects for the decreasing  number of  terms in the denominator 
of the second factor. To obtain identical time bins for both curves,
we linearly interpolated the radio data to the grid defined by the X-ray bins; 
this is sufficiently accurate given the short time bins compared to the intrinsic 
light curve variability time scales. The inset in Figure~\ref{lightcurves}c shows $P(\ell)$.
The function sharply peaks at zero lag, indicating no significant lag between
the two curves. Note that the units of
the X-ray time derivative and the radio flux are not important for equation~\ref{neupert}
and the figures have been scaled arbitrarily. Also, the X-ray time derivative
is mostly negative since the variability occurs during the gradual decline from the 
main peak. Again, this is of little importance as such a trend can be subtracted 
from the data, shifting the derivative to larger values.

\section{Discussion}

Figure~\ref{lightcurves} shows compelling evidence for the presence of 
a Neupert effect, i.e., a radio light curve that is approximately proportional 
to the  time derivative of the X-ray light curve. 
The light curves alone do not prove the operation of chromospheric
evaporation induced by electron beams. It is possible that the rate
of electron injection is proportional to the rate of heating, both being 
induced by the same mechanism,  while
the two processes are causally unrelated. A causal relation is difficult
to demonstrate even in the case of the Sun, although
the evidence for electron beam heating in many solar flares is compelling.

  A causal relation between particle acceleration and coronal
heating is however supported if the energy in the non-thermal electrons
is sufficient to heat the observed plasma. Further, this energy
should not be radiated away by the high-energy particles but be transported 
to the  chromosphere on time scales shorter than the radiative loss
time. An accurate
calculation requires information that is not at hand,  but we present
estimates for a plausibility analysis as follows: By estimating the
gyrosynchrotron radio flux and assuming reasonable magnetic field strengths and electron
distributions, we infer the total kinetic energy in the electrons at
any given time. By integrating the energy across the flare, we can compare the
total injected energy with the total X-ray losses. For such an order-of-magnitude estimate,
we will assume that the shape of the electron distribution stays constant (see below), and
that variable radio self-absorption is not relevant. We perform the estimates
for a range of given but, in each case, constant values of the magnetic field strength and
of the slope of the electron distribution.

Non-thermal electrons in stellar coronae and solar flares are usually
distributed in energy in a power law of the form
\begin{equation}\label{distribution}
n(E,t) = {N(t)(\delta -1)\over E_0} \left({E\over E_0}\right)^{-\delta}
\end{equation}
where $n(E,t)dE$ is the time-dependent number density of electrons in the energy
interval $[E, E+dE]$,  $E_0$ is the lower cutoff energy for the distribution
of non-thermal electrons, $N(t)$ is the normalization constant, viz. the volumetric
number density of electrons at time $t$ above $E_0$, and $\delta > 1$ is the power-law index
(typically $\delta = 3-5$ in solar flares, \citealt{dennis88}).
In solar flares, there is evidence that one and the same electron power-law distribution
comprises the (lower-energy) hard-X-ray emitting
electrons and the (higher-energy) microwave-emitting  electrons \citep*{bastian98}.
The value of $E_0$ is unknown but must be $>0$ in order 
to confine the total non-thermal energy \citep{dennis88}. We set $E_0 = 10$~keV. This 
is likely to be a lower limit for solar flares where $E_0 = 20-30$~keV may
be more appropriate \citep{dennis88}.
 

Radio flare emission from RS CVn binaries at 5~GHz is often optically
thick \citep{mutel85,white95}. Optically thick emission from a simple source 
should stay constant, independent of the number of injected electrons, but this
is not usually observed \citep{morris90}, indicating that the number of electrons 
itself is responsible for continually changing the optically thick cross section, e.g.,
by being injected in a variable number of magnetic loops. We will
thus assume that the volume filled by non-thermal electrons is generally 
time-dependent.

The radio spectral index at optically thick low frequencies is typically +1, 
while the index on the high-frequency optically thin side is between $-$0.5 and $-$1 
\citep{feldman78, morris90, jones96, beasley02}. The turnover frequency
(above which  the radiation changes to optically thin) is similar
in many reported flares, $\nu_{\mathrm{peak}} \approx 10-20$~GHz although
values as low as a few GHz are possible \citep{morris90}. From this typical spectral
model shape, we can estimate the unabsorbed luminosity 
at 5~GHz to be $\sim 1-10$ times higher than measured. 
The unabsorbed radio flux $f_{\mathrm{R}}$ (taken to be 5 times the observed
flux in what follows)  relates to the gyrosynchrotron emissivity $\eta$, for which  
\citet{dulk82} gave the following approximation for the magnetoionic x-mode:
\begin{equation}\label{emissivity}
\eta_x = 3.3\times 10^{-24}10^{-0.52\delta}BN(\mathrm{sin}\theta)^{-0.43+0.65\delta}
\left({\nu\over \nu_{\mathrm{B}}}\right)^{1.22-0.90\delta}
\end{equation}
(in erg~cm$^{-3}$~s$^{-1}$~Hz$^{-1}$~sr$^{-1}$). $B$ (in units of Gauss) is the magnetic 
field strength (assumed to be constant in the source), and $\nu_{\mathrm{B}} \approx 2.8\times 
10^6B$~[Hz] is the electron gyrofrequency.  We will use $\eta = 2\eta_x$
for the total emissivity, assuming a similar emissivity for the o-mode which is 
approximately valid under typical coronal conditions. We note that in general the 
o-mode emissivity is smaller \citep{dulk82}; we thus overestimate $\eta$ somewhat, and
hence we will underestimate the total energy content in the electrons accordingly.
Since most of the emission is radiated  by electrons with large pitch angles,
we set $\theta = 60^{\circ}$ in what follows. This represents an average for a  
uniform pitch angle distribution with $\delta$ around 3 in equation~\ref{emissivity}. 
The radio flux then is
\begin{equation}\label{luminosity}
f_{\mathrm{R}}(t) = {\eta V(t)\over d^2} = {f(B,\delta,\nu)\over d^2}N(t)V(t)
\end{equation}
where $V$ is the source volume, $d$ is the distance to the source, and 
$f(B,\delta,\nu)$ collects various terms from equation~\ref{emissivity}. The total 
instantaneous kinetic energy in the electron distribution (given in equation \ref{distribution}) 
is
\begin{eqnarray}\label{energy}
E_{\mathrm{kin}}(t) &=& N(t)V(t)(\delta - 1)E_0^{\delta - 1}\int_{E_0}^{\infty}E^{-(\delta-1)}dE 
\nonumber \\
 &=&    N(t)V(t){\delta - 1 \over \delta - 2}E_0
\end{eqnarray}
where $\delta > 2$ has been assumed.
The expression $N(t)V(t)$ is obtained from equation~\ref{luminosity} with the 
known radio flux $f_{\mathrm{R}}$. We finally need to specify values for $\delta$ and 
$B$. Our single-frequency observation is not sufficient to derive 
these values. However, from a large body of published modeling results,
likely values are within the ranges $2.0 < \delta \le 3.5$ and 
$20 \le B \le 200$~G (e.g., \citealt{mutel85, morris90, chiuderidrago93, jones94}).
We consider the range $2.1 \le \delta \le 3.5$.  

Travel times of electrons of 10$-$100~keV along a magnetic loop
of $10^9$~cm are of order 1~s, i.e., if the electrons all get lost in the 
deeper atmospheric layers, then  equation~\ref{energy} gives the energy that
should be replaced every second. However, radio-emitting electrons
are probably trapped in magnetic fields for some time by virtue of their
large pitch angles, while a large fraction of the lower-energy electrons that
carry most of the energy to heat the chromosphere arrive there earlier.

The decay time $\tau$ cannot exceed the 
synchrotron loss time $\tau_{\mathrm{s}}$,
\begin{equation}\label{taus}
\tau_s = {6.8\times 10^8\over B^2\gamma}\quad \mathrm{[s]}
\end{equation}
\citep{petrosian85}, where $\gamma$ is the Lorentz factor of the electrons. 
For electrons  observed at a frequency $\nu$,
\begin{equation}
\nu \approx \gamma^2\nu_{\mathrm{B}}.
\end{equation}
Substituting $\nu_{\mathrm{B}}$ and inserting $\gamma$ into equation~\ref{taus} leads to
\begin{equation}
\tau_{\mathrm{s}} = {1.1\times 10^{12}\over B^{3/2}\nu^{1/2}}\quad \mathrm{[s]}.
\end{equation}
For our ranges of $B$, $\tau_s$ is minimized  for 
$B = 200$~G, $\tau_{\mathrm{s}} = 5500$~s. However, the fastest
observed decay times in the radio light curve are of order 
2000~s and possibly as small as 1000~s. We conclude that, in the framework of our
simplistic model,   losses other than
synchrotron losses (most likely scattering into the loss cone and precipitation
into the chromosphere, possibly also collisional losses) predominate.
We henceforth adopt an upper limit to the lifetime of radio-emitting
electrons of $\tau = 1500$~s.

The total energy injection rate is thus overestimated from the radio information. 
Starting at time $t = 0$ and assuming that the complete particle distribution `decays'
independently of energy by continuously losing energy (by various processes as mentioned above)
and thus eventually thermalizing, the  instantaneous number of radio-emitting particles evolves as
\begin{equation}\label{evolution}
N(t)V(t) = N(0)V(0)e^{-t/\tau} + \int_0^t{d[N_+(t^{\prime})V(t^{\prime})])\over dt^{\prime}}e^{-(t-t^{\prime})/\tau}dt^{\prime} 
\end{equation}
where $d[N_+V]/dt \ge 0$ is the injection rate of new electrons.
The first term on the right-hand side describes the decaying density of 
particles already present at $t=0$, while the second term describes the evolution
of newly injected electrons. 

We note that equation~\ref{evolution}
implies a convolution of the electron injection rate with a cut-off exponential
function. The injection rate should thus peak considerably before the 
observed peak of $f_{\mathrm{R}} \propto N(t)V(t)$ if $\tau$ were large. We take the very close temporal
coincidence between the radio light curve and the derivative of the 
X-ray light curve as further a-posteriori support for a small energy loss
time for the high-energy electrons. For $\tau = 1000$~s, the time delay
between peak injection rate and  peak $N(t)V(t)$ is negligible, while for 
 $\tau = 1500$~s, it amounts to about 1000~s, which is still rather small.  
Equation~\ref{evolution} implies
\begin{equation}\label{rate}
{d[N_+(t)V(t)]\over dt} = {d[N(t)V(t)]\over dt} + {N(t)V(t)\over \tau}.
\end{equation}
In what follows, we subtract the apparently quiescent emission level of 3~mJy from
the radio light curve (Figure 1c). The expression involving $N(0)$ in 
equation~\ref{evolution} can therefore be neglected, also because
of the rather rapid decay time. It is easy to show that the rate of
electron injection $d[N_+V]/dt$ is positive at a given time $t$ if the observed 
decay time scale of $f_{\mathrm{R}}(t) \propto N(t)V(t)$ is larger than the assumed intrinsic 
decay time $\tau$ of the electron population. Since the latter is a necessary condition,
we fulfil $d[N_+V]/dt > 0$ by adopting a value for $\tau$ at the lower limit
of the measured decay time scales in the radio light curve, as done above.

\begin{figure}
\epsscale{1.}
\plotone{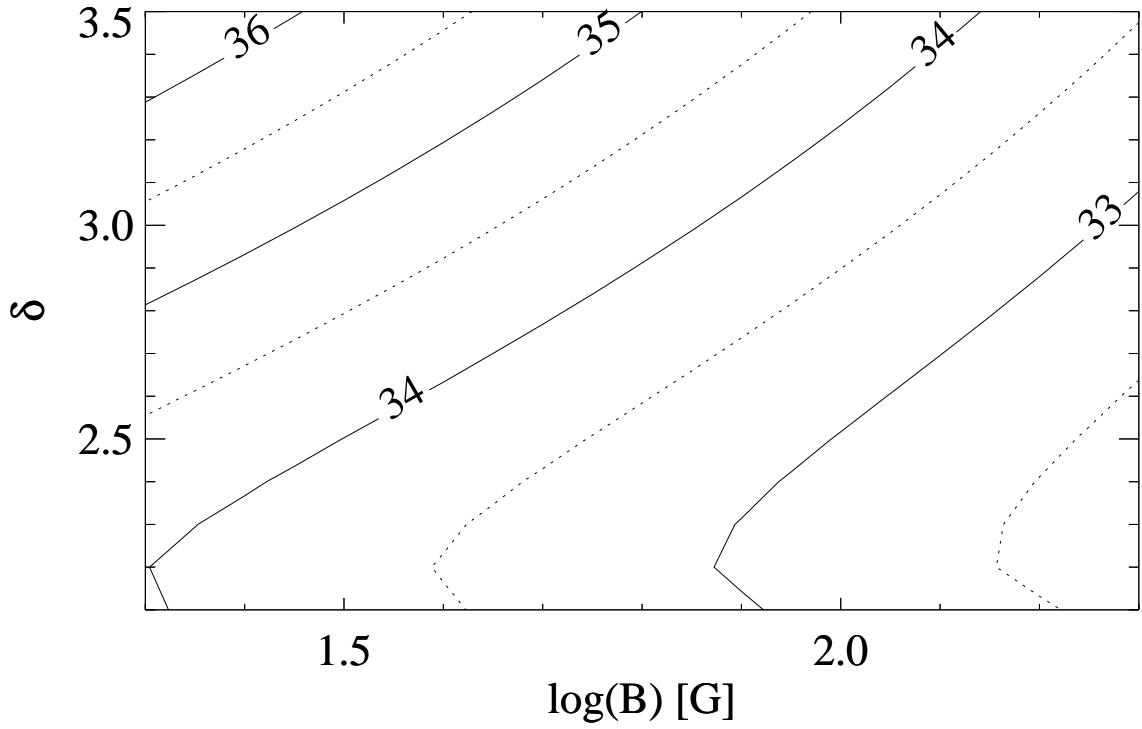}
\caption{Estimated kinetic energies $E$ in accelerated electrons (above 10~keV), integrated
       over the complete second radio flare, as a function of the magnetic field strength and
       the power-law index $\delta$.  Solid contours are labeled with log$E$, and dashed
       contours refer to half-decade log$E$ values (log$E$ = 32.5 etc).
 \label{contour}}
\end{figure}

When the rate of change in the derived instantaneous energetic-particle number $N(t)V(t)$
is small (first term on the right-hand side of equation~\ref{rate}, e.g., around the radio peak time), 
then the injection rate of particles is equal to $N(t)V(t)/\tau$, and the injected
power analogously equals the total energy content divided by  $\tau$. The total 
number of injected electrons can be obtained by integrating equation~\ref{rate}
in time across the complete radio flare (time interval 
[$T_0$,$T$] = [1.03,1.32]~d). Since the radio light curve  approximately returns 
to the pre-flare level, $N(t)V(t)$ in equation~\ref{luminosity} returns to the same 
value; thus, the first term in equation~\ref{rate} does not contribute 
to the integral, and the total number of injected electrons is
\begin{eqnarray}\label{integral}
{\cal{N}}&=& {1\over \tau}\int_{T_0}^TN(t)V(t)dt  \nonumber \\
     &=& 1.55\times 10^{19}10^{3.49\delta}B^{0.22-0.90\delta}
        {d^2\over \tau}\int_{T_0}^Tf_{\mathrm{R}}(t)dt.
\end{eqnarray}
The total injected energy can then
be obtained from ${\cal{E}} = {\cal{N}}E_0(\delta-1)/(\delta-2)$ (equivalent to 
equation~\ref{energy}). We have
performed the integration~\ref{integral} for the $B$ and $\delta$ values
allowed here. The result is graphically shown in Figure~\ref{contour}.
For the selected parameter range, the total injected energy is of order
$10^{33} - 10^{36}$~erg. We compare this energy with the total energy loss
in the second X-ray flare after 1.03~d. Judged from the count rate, the excess 
luminosity above the slow decay from the first flare peak
reaches $\approx 3\times 10^{30}$~erg~s$^{-1}$. With a half-duration of about 0.15~d,
we obtain a total radiated energy of $4\times 10^{34}$~erg. This 
value compares favorably with the estimates in Figure~\ref{contour}.
While there is no proof for a causal relation between high-energy
electrons and coronal heating, the energy budget is compatible with 
such a scenario for a reasonable range of parameters $B, \delta$.

\section{Conclusions}


We note that, within the framework of our simplified model assumptions,  our 
energy estimate is conservative. The lifetime $\tau$ of the non-thermal
electron population has been adopted at the highest possible value compatible 
with the radio light curve (i.e., approximately equal to the shortest
decay time scale in the observed radio emission). Equation~\ref{integral} shows that a shorter
possible lifetime requires a proportionately larger total energy input. In that case, the
light curve variability is controlled by the time scales of the particle accelerator.
If electrons get lost after one magnetic-loop crossing time, then $\tau \approx$ a few 
seconds for typical loop sizes, and the rapid replenishment of electrons requires an energy 
input into the system up to 3 orders of magnitude higher than estimated above. We conclude
that not only does the relative timing between radio and X-ray emissions support the
chromospheric evaporation scenario, but the total energy content in the
injected high-energy electrons could easily satisfy or largely exceed the 
requirements set by the observed X-ray losses. 

The observation described here provides strong support for the chromospheric evaporation
scenario in a star that may maintain a corona considerably different from the
Sun's (e.g., containing much larger magnetic loops, confining much hotter
and perhaps also denser thermal plasma, magnetic field lines that may be arranged
in the form of large global dipoles, possibly also between the companion stars, etc).
In retrospect, we find a similar timing between radio and X-ray flare events in
some previously published light curves, although the Neupert effect was not discussed.
Most evidently, radio emission peaking before the soft X-rays, thus suggesting
the presence of a Neupert effect,  can be seen in the examples presented by
\citet{vilhu88}, \citet{stern92},  \citet{brown98}, and \citet{ayres01}.
Clearly, differing behavior has been noted as well. First, the Sun shows the 
Neupert effect most reliably in the class of impulsive flares, whereas 50\% of all
gradual flares, often related to energy release at high coronal altitudes, 
show a different behavior \citep{dennis93}. It is possible that 
the connectivity of magnetic fields between the high corona and the chromospheric
regions is different in these cases, impeding the free flow of electrons
and consequent chromospheric evaporation. Thermalization of fast electrons could 
also occur in the corona already if the travel distances are long enough and the
densities high enough. Stellar counter-examples of the Neupert effect include an impulsive optical flare 
with following gradual radio emission \citep{vdoord96}, gyrosynchrotron emission that 
peaks after the soft X-rays \citep{osten00}, and an X-ray depression during strong radio flaring
\citep{guedel98}. Note also that complete absence of correlated flaring has been observed 
at radio and UV wavelengths (e.g., \citealt{lang88}).

Evidence for  chromospheric evaporation in an RS CVn binary system is potentially
important to understand to what degree the solar analogy can be applied to such 
stellar systems. Although magnetospheric sizes as measured in radio waves are
several times larger than the Sun, VLBI observations have suggested that
flares start out in compact, unresolved cores that are well localized in 
magnetic active regions close to the stellar surface \citep{mutel85}. Our observations of
a Neupert effect strongly suggest solar analogy in the physics of energy release
and transport in this binary system, at least for the large flare reported here.


\acknowledgments  We acknowledge the hospitality of CASA/University of Colorado
                  where most of this paper was written. 
		  Research at PSI  has been supported by the Swiss National 
                  Science Foundation (grant 2100-049343). SRON is supported financially 
		  by NWO. The present project is based 
                  on observations obtained with {\it XMM-Newton}, an ESA science 
                  mission with instruments and contributions directly funded by 
                  ESA Member States and the USA (NASA). The VLA is a facility of the 
		  National Radio Astronomy Observatory, which is operated by Associated 
		  Universities, Inc., under cooperative agreement with the  
		  National Science Foundation.
 

\clearpage

\begin{thebibliography}{}
\bibitem[Ayres et al.(2001)]{ayres01}Ayres, T.~R., Brown, A., Osten, R.~A., Huenemoerder, D.~P.,
          Drake, J.~J., et al. 2001, ApJ, 549, 554 
\bibitem[Bastian et al.(1998)Bastian, Benz, \& Gary]{bastian98}Bastian, T.~S., 
         Benz, A.~O., \& Gary, D.~E. 1998, ARA\&A, 36, 131 
\bibitem[Beasley et al.(2002)]{beasley02}Beasley, A.~J., et al. 2002, in preparation
\bibitem[Brown et al.(1998)]{brown98}Brown, A., Osten, R.~A., Drake, S.~A., Jones, K.~L.,
         \& Stern, R.~A. 1998, In The Hot Universe. IAU Symp 188,  
	 Eds. K Koyama, S. Kitamoto, \& M. Itoh (Dordrecht: Kluwer), 215 
\bibitem[Chiuderi Drago \& Franciosini(1993)]{chiuderidrago93}Chiuderi Drago, F.,
          \& Franciosini, E. 1993,  ApJ, 410, 301 
\bibitem[Dennis(1988)]{dennis88}Dennis, B.~R. 1988, Solar Phys.,  118, 49 
\bibitem[Dennis \& Zarro(1993)]{dennis93}Dennis, B.~R., \& Zarro, D.~M. 1993, 
         Solar Phys., 146, 177 
\bibitem[Drake et al.(1989)Drake, Simon, \& Linsky]{drake89}Drake, S.~A., Simon, T., 
         \& Linsky, J.~L. 1989,  ApJ, 71, 905 
\bibitem[Dulk \& Marsh(1982)]{dulk82}Dulk, G.~A., \& Marsh, K.~A. 1982,  ApJ,
            259, 350 
\bibitem[Engvold et al.(1988)]{engvold88}Engvold, O., et al. 
         1988,  A\&A, 192, 234
\bibitem[ESA(1997)]{esa97}ESA 1997, The Hippacos and Tycho Catalogues, ESA SP-1200
\bibitem[Feldman et al.(1978)]{feldman78}Feldman, P.~A., Taylor, A.~R., Gregory, P.~C.,
          Seaquist, E.~R., Balonek, T.~J., Cohen, N.~L. 1978, AJ, 83, 1471
\bibitem[G\"udel et al.(1996)]{guedel96}G\"udel, M., Benz, A.~O., Schmitt, J.~H.~M.~M.,
        \& Skinner S.~L.  1996, ApJ, 471, 1002
\bibitem[G\"udel et al.(1998)]{guedel98}G\"udel, M., Guinan, E.~F., Etzel, P.~B., Mewe, R.,
          Kaastra, J.~S., \& Skinner, S.~L. 1998, 
	 In 10th Cambridge Workshop on Cool Stars, Stellar Systems, and the Sun,
         ed. R. Donahue \& J.~A Bookbinder, (San Francisco: ASP), 1247 
\bibitem[Hawley et al.(1995)]{hawley95}Hawley, S.~L., et al. 1995, ApJ, 453, 464 
\bibitem[Hudson \& Ryan(1995)]{hudson95}Hudson, H., \& Ryan, J. 1995, ARA\&A, 33, 239
\bibitem[Jansen et al.(2001)]{jansen01}Jansen, F., et al. 2001, A\&A, 365, L1
\bibitem[Jones et al.(1994)]{jones94}Jones, K.~L., Stewart, R.~T., Nelson, G.~J., 
          Duncan, A.~R. 1994,   MNRAS, 269, 1145 
\bibitem[Jones et al.(1996)]{jones96}Jones, K.~L., Brown, A., Stewart, R.~T., \& Slee, O.~B. 
         1996, MNRAS, 283, 1331 
\bibitem[Lang \& Willson(1988)]{lang88}Lang, K.~R., \& Willson, R.~F. 1988, ApJ, 328, 610
\bibitem[Morris et al.(1990)Morris, Mutel, \& Su]{morris90}Morris, D.~H., Mutel, R.~L., \&
        Su, B. 1990,  ApJ, 362, 299
\bibitem[Mutel et al.(1985)]{mutel85}Mutel, R.~L., Lestrade, J.-F.,  Preston, R.~A., 
         \& Phillips, R.~B. 1985,  ApJ, 289, 262
\bibitem[Neupert(1968)]{neupert68}Neupert, W.~M. 1968, ApJ, 153, L59 
\bibitem[Osten \& Brown(1999)]{osten99}Osten, R.~A., \& Brown, A. 1999, ApJ, 515, 746 
\bibitem[Osten et al.(2000)]{osten00}Osten, R.~A., Brown, A., Ayres, T.~R.,
         Linsky, J.~L., Drake, S.~A., Gagn\'e, M.,  Stern, R.~A. 2000, ApJ, 544, 953 
\bibitem[Pallavicini, Willson, \& Lang(1985)]{pallavicini85}
          Pallavicini, R., Willson, R.~F., \& Lang, K.~R. 1985, A\&A, 149, 95
\bibitem[Petrosian(1985)]{petrosian85}Petrosian, V. 1985,  ApJ, 299, 887
\bibitem[Sakao(1994)]{sakao94}Sakao, T. 1994, Characteristics of solar flare hard
         X-ray sources as revealed with the hard X-ray telescope aboard the Yohkoh satellite.
	 PhD thesis. (Tokyo: University of Tokyo).
\bibitem[Schrijver et al.(1995)]{schrijver95}Schrijver, C.~J., Mewe, R., van den Oord, 
         G.~H.~J., \& Kaastra, J.~S. 1995, A\&A, 302, 438
\bibitem[Singh et al.(1987)]{singh87}Singh, K.~P., Slijkhuis, S., Westergaard, N.~J.,
           Schnopper, H.~W., Elgaroy, O., Engvold, O., \& Joras, P. 1987,  MNRAS, 224, 481
\bibitem[Stern et al.(1992)]{stern92}Stern, R.~A., Uchida, Y., Walter, F.~M., Vilhu, O.,
         Hannikainen, D., Brown, A., Veal\'e, A., \& Haisch, B.~M. 1992, ApJ, 391, 760 
\bibitem[Strassmeier et al.(1993)]{strassmeier93}Strassmeier, K.~G., Hall, D.~S.,
         Fekel, F.~C., \& Scheck, M. 1993, A\&AS, 100, 173
\bibitem[Str\"uder et al.(2001)]{strueder01}Str\"uder, L., et al. 2001, A\&A, 365, L7
\bibitem[van den Oord et al. (1996)]{vdoord96}van den Oord, G.~H.~J., Doyle, J.~G., 
        Rodon\`o, M., Gary, D.~E., Henry, G.~W., et al.  1996, A\&A, 310, 908
\bibitem[Vilhu et al.(1988)]{vilhu88}Vilhu, O., Caillault, J.-P., \& Heise, J. 1988,
           ApJ, 330, 922 
\bibitem[White \& Franciosini(1995)]{white95}White, S.~M., \& Franciosini, E. 1995,
         ApJ, 444, 342
\bibitem[Yi et al.(1997)]{yi97}Yi, Z., Elgaroy, O., Engvold, O., \& Westergaard, N.~J.
         1997, A\&A, 318, 791
\end{thebibliography}
\end{document}